\documentclass[usenatbib]{mnras}
\usepackage{epsfig}
\usepackage{amsmath,amssymb}
\usepackage{color}
\usepackage{booktabs}
\usepackage{pdflscape}
\usepackage{textcomp}

\usepackage{epsfig}
\usepackage{color}
\usepackage{soul}
\usepackage{aas_macros}  
\usepackage{chngcntr}
\graphicspath{{figures/}}

\newcommand{\Rsolar}{\mbox{\,$\rm R_{\odot}$}}        

\newcommand{\Rstar}{\mbox{\,$\rm R_{\star}$}}        

  \newcommand{\cmss}{\,\mbox{$\mbox{cm}\,\mbox{s}^{-2}$}}    
  \def\simge{\mathrel{\raise1.16pt\hbox{$>$}\kern-7.0pt
    \lower3.06pt\hbox{{$\scriptstyle \sim$}}}}           
  \def\simle{\mathrel{\raise1.16pt\hbox{$<$}\kern-7.0pt
    \lower3.06pt\hbox{{$\scriptstyle \sim$}}}}           

\newcommand{\perday}{\mbox{\,$\rm d^{-1}$}}        

\title[Photometry of V348\,Sgr]{ROAD\footnote{Remote Observatory Atacama Desert} and K2 photometry of V348\,Sgr: probing the pulsation dust connection}

\author[C. S. Jeffery \& F.-J. Hambsch]{C. S. Jeffery$^{1,2,3}$\thanks{email: Simon.Jeffery@armagh.ac.uk} and F.-J. Hambsch$^{4,5,6}$\\
$^{1}$ Armagh Observatory and Planetarium, College Hill, Armagh BT61 9DG, UK\\
$^{2}$ School of Physics, Trinity College Dublin, College Green, Dublin 2, Ireland\\
$^{3}$ Institute of Astronomy, University of Cambridge, Cambridge CB3 0HA, UK\\
$^{4}$ Vereniging Voor Sterrenkunde (VVS), Brugge, Belgium\\
$^{5}$ Bundesdeutsche Arbeitsgemeinschaft f\"ur Ver\"anderliche Sterne e.V. (BAV), Berlin, Germany\\
$^{6}$ American Association of Variable Star Observers (AAVSO), Cambridge, USA\\
}
\begin{document}

\date{Accepted \ldots. Received \ldots; in original form \ldots}

\pagerange{\pageref{firstpage}--\pageref{lastpage}} \pubyear{2014}

\maketitle

\label{firstpage}

\begin{abstract}
V348\,Sgr is simultaneously an active hot R\,Coronae Borealis (RCB) variable, a peculiar extreme helium star, and the hydrogen-deficient central star of a planetary nebula. 
Explaining the RCB-type variability has been difficult since the star spends much of its time at minimum light. 
We present new ground-based multicolour photometry covering 5 observing seasons and 80 days of continuous photometry from space. 
The latter demonstrate small-amplitude ($<0.01$ mag.) variability at maximum light on timescales typical for strange-mode pulsation in hot helium supergiants. 
These could provide a trigger for frequent dust-production episodes; other mechanisms must also be considered.
Multi-colour photometry probes the reddening properties of extinction events from minimum to maximum light. 
The latter are comparable with extinction events due to carbonaceous grains seen in cooler RCB stars. 
Minimal reddening at minimum light is indicative that starlight scattered from circumstellar dust into the line of sight dominates transmitted light.
\end{abstract}


\section{Introduction}
\label{s:intro}
V348\,Sgr (= HV 3976 = 21.1929 Sgr) \citep{herbig58} 
has been described as a Rosetta stone for stellar evolution \citep{schoenberner86cx}. 
It is at the same time a peculiar extreme helium star\footnote{ \citet{leuenhagen94} describe V348\,Sgr as a low-mass carbon-rich Wolf-Rayet star of spectral type [WC12].  \citet{crowther98} observe that V348\,Sgr fails their criteria for a [WCL] classification and prefer `peculiar extreme helium star', a designation followed by \citet{acker03}.},
the central star of a planetary nebula (PN G011.1-07.9 = SB17) \citep{herbig58,pollacco91b},
and is host to an extended dust-disk and envelope \citep{clayton11b}.
It shows light variations of the R\,Coronae Borealis (RCB) type over timescales of months \citep{woods26,schajn29,parenago31,hoffleit58,bateson82},
low hydrogen and high carbon abundance \citep{jeffery95b,leuenhagen94,leuenhagen96} 
and a secular fading over a timescale of years which probably corresponds to an increase in effective temperature as the star contracts \citep{demarco02,schaefer16}. 

Analyses of its photosphere and wind find V348\,Sgr to have a high luminosity-to-mass ratio, an effective temperature 20\,000 -- 22\,000\,K, surface gravity $\log g/\cmss < 2.7$, radius 6\Rsolar, a surface characterized principally by mass fractions of $<4$\% hydrogen, and approximately 40\% helium and 55\% carbon, and a radiatively-driven wind indicative of mass loss of some 
$\dot{M}\approx 10^{-6.5} {\rm M_{\odot} yr^{-1}}$ \citep{dahari84,jeffery95b,leuenhagen94,leuenhagen96}.

RCB-type extinction events are associated with the ejection of dusty material. 
Multi-wavelength studies of the circumstellar dust around V348\,Sgr have demonstrated its gross properties by comparing spectra obtained at light maximum and light minimum,  concluding that its composition is dominated by amorphous or disordered carbon grains \citep{feast73,schoenberner86bx,jeffery95a,drilling97,hecht98,wada98,clayton11b,gavilan17}. 

Whilst \citet{herbig58} speculated that the natural state of V348\,Sgr is at minimum light interrupted by luminous outbursts, the contemporary view is that a naturally  bright state is interrupted by the ejection of carbon-rich gas which condenses to form dust that obscures some or all of the stellar surface.
Explanations for the surface chemistry and the presence of a nebula are provided by a star which after evolving through its hydrogen and helium-burning phases has left the asymptotic giant-branch and which experienced a late thermal pulse (or helium shell flash) as it was approaching the white dwarf cooling track. 
This pulse caused the star to expand rapidly and to become briefly a red giant and now to contract toward the white dwarf sequence for a second time \citep{dahari84,schoenberner86ax,pollacco90,herwig00,demarco02}. 
This history is distinct from that  attributed to cooler RCB and extreme helium (EHe) stars, believed to arise from helium-shell burning giants which form following the merger of two white dwarfs \citep{webbink84,iben84,saio00,saio02,clayton07,jeffery11a,zhang13,zhang14}. 

V348\,Sgr continues to pose questions. 
There has been no study of the evolution of dust properties during a V348\,Sgr extinction event,
nor has it been established whether these events and those seen in the classical (cooler) RCB stars are equivalent. 
Most cool RCB stars spend less than 20\% of their time at minimum light; 
dust formation occurs at a radial distance of $\approx2\Rstar$ on timescales short ($\leq 1$ week) compared with the dynamical timescale ($\approx40$\,d).
In contrast, the temperature and ultraviolet flux at 2\Rstar\ from V348\,Sgr are too high for dust nucleation to occur,  
and V348\,Sgr spends up to 45\% of time at minimum light \citep{hoffleit58,bateson82}. 
The dust formation timescale is long compared with the dynamical time. 
Ultimately it will be necessary to establish how far the dust forms from the star, what is the chemistry and nucleation mechanism that triggers dust formation, and what trigger can be so efficient to maintain the observed level of activity. 
Resolving these questions has hitherto been hindered by a scarcity of synoptic photometry at any wavelength, let alone in more than one waveband.

This paper  presents results from a continuing campaign to monitor V348\,Sgr in three wavebands, and from the {\it Kepler} spacecraft which observed V348\,Sgr during {\it K2} Campaign 7 (\S\,2). 
Photometry at maximum light is used to look for short-period surface activity (\S\,3). The connection between surface activity and extinction  triggers is discussed in conjunction with the reddening properties of dust derived from three-colour photometry (\S\,4). 
Corollaries between extinction events in V348\,Sgr and R\,CrB are discussed.

\begin{figure}
\begin{center}
    \epsfig{file=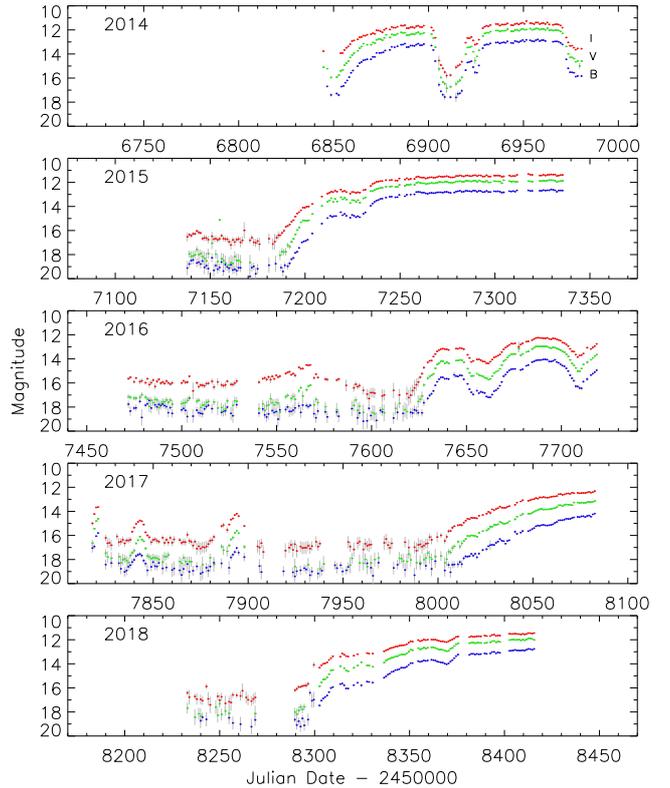,width=85mm,clip=,angle=0}
\end{center}
\caption{Three-colour ($BVI$) photometry of V348\,Sgr obtained at ROAD between 2014 and 2018.   } 
\label{f:v348_lc2}
\end{figure}

\begin{figure}
\begin{center}
    \epsfig{file=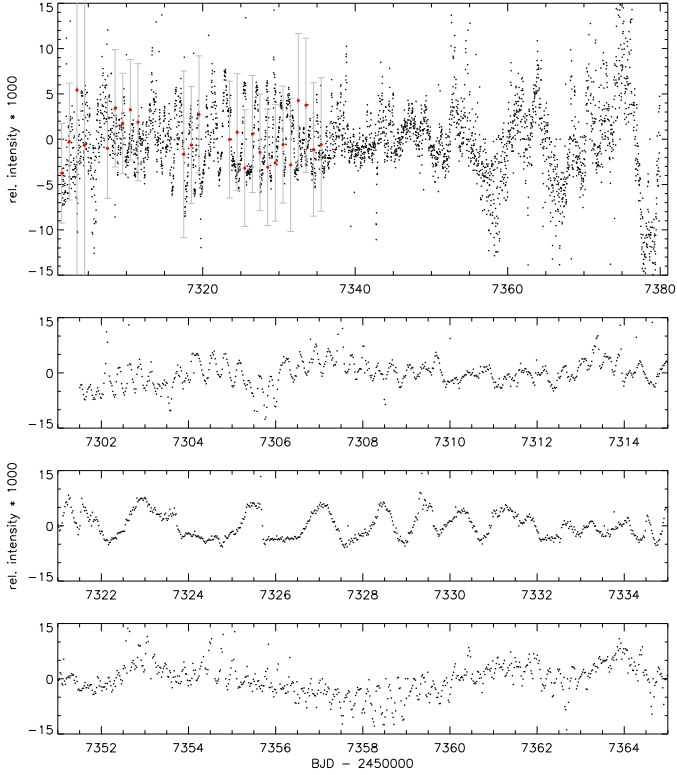,width=90mm,clip=,angle=0}
\end{center}
\caption{{\it K2} photometry of V348\,Sgr obtained between 2015 Oct 04 and 2015 Dec 26 shown as difference relative to the local mean flux. The three lower panels enlarge selected segments of the complete light curve shown in the top panel. The contemporary ROAD photometry ($I$) is shown in the top panel (large red dots with error bars). } 
\label{f:v348_k2}
\end{figure}

\section{Observations}
\label{s:stability}

V348\,Sgr varies with a range from 10.6 to 17.1 in photographic magnitude, spending more time at minimum light than at maximum \citep{hoffleit58}. 
\citet{heck85} state that  ``Additional photometric observations are necessary to \ldots gain insight into the fading and brightening phases [of V348\,Sgr]. As it is not possible to mobilize professional instruments for monitoring this unpredictable star, we have here a typical case where amateur astronomers might be of invaluable assistance''. 
With the increasing capability of amateur astronomers and their observatories, V348\,Sgr offers a promising target. 

\begin{table}
    \caption{Comparison stars used for differential photometry of V348\,Sgr}
    \centering
    \setlength{\tabcolsep}{1.5pt}
    \begin{tabular}{llllll}
 ID & $\alpha(2000)$ & $\delta(2000$ & $V$ & $I$  \\    
R: UCAC4 335-181356	& 18:39:32.65	& --23:07:13.50 & 11.29 & 11.20 \\
Ck: UCAC4 336-170188	& 18:40:23.47	& --22:50:04.14 & 11.74 & 11.17 \\
C1: 000-BCC-932	 	& 18:40:05.27	& --22:51:19.8 & 12.96 & 12.37 \\
C2: 000-BCC-939	 	& 18:40:14.94	& --22:54:03 & 14.40 & 12.97 \\
    \end{tabular}
    \label{t:comp}
\end{table}

\subsection{Remote Observatory Atacama Desert Observations}

The Remote Observatory Atacama Desert (ROAD) is located at San Pedro de Atacama, Chile at an  altitude of 2400\,m and with an average of 320 clear nights per year. 
It is equipped with a 40 cm f/6.8 Optimized Dall Kirkham reflector from Orion Optics, UK,
a ML16803 CCD camera (FLI, USA) and Astrodon $BVI$ photometric filters \citep{hambsch12a}.

Observations of V348\,Sgr commenced 2014 August. 
Near nightly $BVI$ observations  were obtained during the observing seasons  from 2014 to 2018.
Data were reduced using the LesvePhotometry package\footnote{\tt http://www.dppobservatory.net/AstroPrograms/\\ Software4VSObservers.php}. 
Magnitudes are obtained corrected to a reference star (R). Lightcurves for a check star (Ck) and two other comparison stars (C1, C2) were also extracted (Table\,\ref{t:comp}). 
All data are available online from the AAVSO database (observer code HMB).

The complete $BVI$ light curve for 2014 to 2018 is shown in Fig.~\ref{f:v348_lc2}. 

\subsection{K2 (Kepler) Observations}

Long cadence (LC) observations of V348\,Sgr (=\,EPIC\,216129500) were obtained during Campaign 7 of the {\it Kepler/K2} mission between 2015 Oct 04 and 2015 Dec 26. 
Data were downloaded from the Mikulski Archive for Space Telescopes after reduction with the EVEREST pipeline \citep{everest16}. 
The detrended lightcurve is shown in Fig.~\ref{f:v348_k2}. 
Substantial variations with semi-amplitude between 5 and 10 parts per thousand (ppt) on multiple timescales between 0.5 and 2 days  are evident. 

\begin{figure}
\begin{center}
    \epsfig{file=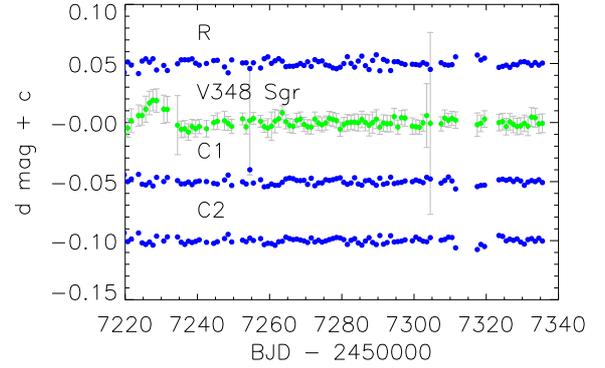,width=80mm,clip=,angle=0}
\end{center}
\caption{ROAD Differential $I$-band light curve segments for V348\,Sgr and three comparison stars. The comparison light curves are offset by integer multiples of 0.05 mag.} 
\label{f:diff}
\end{figure}

\begin{figure}
\begin{center}
    \epsfig{file=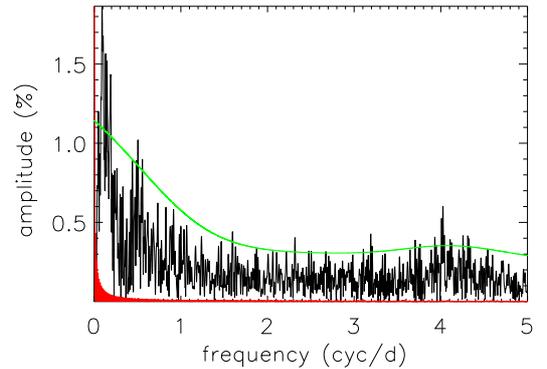,width=80mm,clip=,angle=0}
\end{center}
\caption{Fourier transform periodogram for the {\it K2} light curve of V348\,Sgr  shown in Fig.\,\ref{f:v348_k2}. The window function is shown in red. The smooth (green) curve represents a $4-\sigma$ threshold, indicative of potential periodic signals.} 
\label{f:k2ft}
\end{figure}

\begin{figure*}
\begin{center}
    \epsfig{file=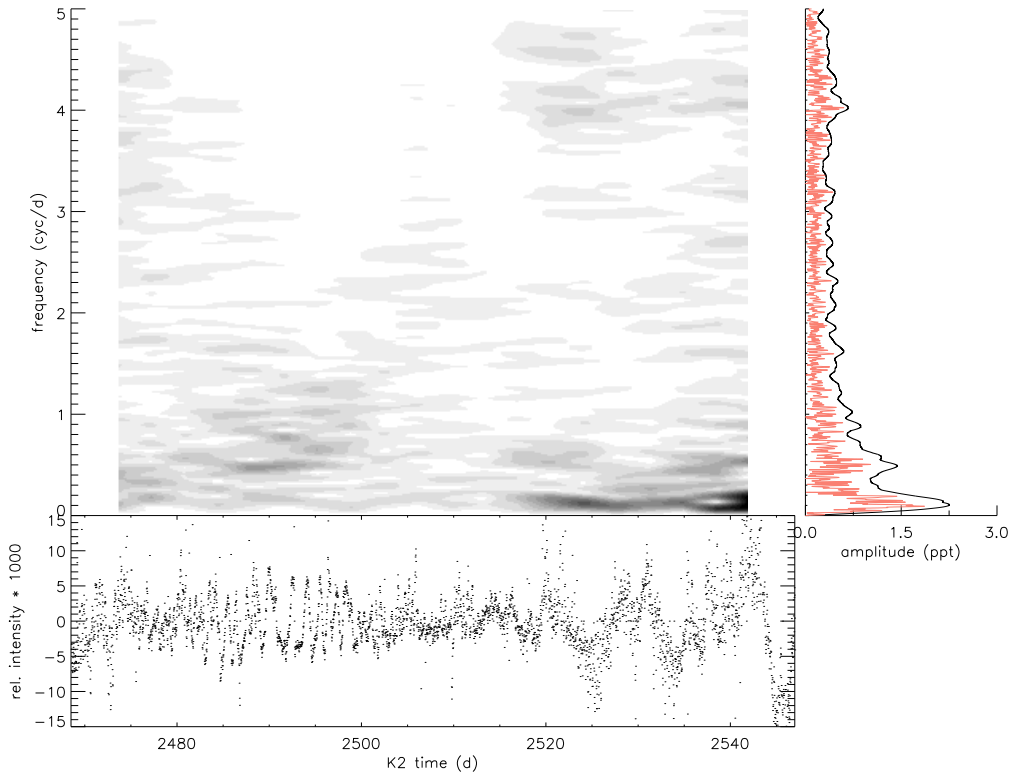,width=140mm,clip=,angle=0}
\end{center}
\caption{Main panel: the sliding amplitude spectrum of the {\it K2}  light curve of V348\,Sgr shown as greyscale, based on 
 data blocks of duration $\Delta T=10$\,d sampled approximately every 5\,d.  The frequency resolution ($1/\Delta T$) 
is hence $\pm0.2\perday$.  The light curve is reproduced at the same horizontal scale in the panel beneath. The panel on the right shows, in bold, the time-averaged amplitude spectrum,  and in light red, the amplitude spectrum of the entire dataset. } 
\label{f:slide}
\end{figure*}

\section{Variability at maximum light}
\label{s:vars}

\subsection{ROAD photometry}

The light curve shown in Fig.~\ref{f:v348_lc2} represents the most complete and only multicolour lightcurve of V348\,Sgr from a single site. 
Whilst the S/N is limited at minimum light, the general form is unsurprising in the context of substantial published records (see \S\,1). 
Approximately 40\% is at light minimum ($V>14$), approximately 25\% at light maximum ($V<11$), with the remainder mostly obtained during egress from minimum or during weak minima.
Unfortunately, the only ingress events captured were not prior to long-lasting deep minima.  

Of the periods at maximum light, an uninterrupted run of some 80\,d in 2015 provides an opportunity to explore the underlying variability at a level of 0.01 mag\footnote{These data were also used by \citet{schaefer16} to point out that the lightcurve only approaches maximum asymptotically, and as evidence that V348\,Sgr has been fading over the past century (at a rate of 1.3 mag century$^{-1}$).}. 
A segment of the I light curve from JD 2457220 to 2457340 has been isolated and inspected. 
The I band has been selected for having the best photon statistics for V348\,Sgr and the reference stars.  The star labelled 'Ck' in Table \ref{t:comp} was used to form differential magnitudes, and corrected to ensure the star labelled 'R' had $\langle I \rangle=11.200$. The lightcurve was detrended using a high-pass Gaussian filter having full width half maximum of 11.8\,d; comparison stars were treated in an identical manner.  The resulting light curves are shown in Fig.\,\ref{f:diff}. 

Excluding an unexplained brightening around JD 2457230, the standard deviations of the measurements of V348\,Sgr and three comparison stars (R, C1, and C2) were 0.0031, 0.0031, 0.0025 and 0.0024 mag. respectively (JD 2457235--2457340).
The mean error on the V348\,Sgr measurements was 0.0083 mag. 
Given these statistics, the appearance of a weak sinusoidal oscillation between JD 2457250 and 2457280 is probably deceptive and we find no unambiguous evidence for short-term small-amplitude variability in the ROAD photometry. 

\subsection{K2 photometry}

The {\it K2} light curve is remarkable. The first quarter of the light curve shown 
in Fig.~\ref{f:v348_k2} (second panel) suggests a period around 0.5\,d or less. 
This half-day period reappears at late times within a more dominant 
variation covering some 5 to 10\,d, but is replaced by  an oscillation of between 1 and 2 days 
during the second quarter. 

An inspection of the  Fourier transform (FT) amplitude spectrum obtained from 
the entire {\it K2} light curve is  equally  perplexing (Fig.~\ref{f:k2ft}). 
It shows two groups of peaks with frequencies $f$ around $0.2$ and $0.4\,{\rm d^{-1}}$, 
and significant isolated peaks at 1.6, 2.4 and $3.2\,{\rm d^{-1}}$ are also evident. 
Power associated with spacecraft pointing corrections is seen at $f=4\,{\rm d^{-1}}$.
Evidently the light variations in V348\,Sgr are non-uniform, varying in both amplitude and frequency.
There may also be multiple signals present which are unresolved in the relatively short duration of 
the {\it K2} campaign. Other  interpretations are also possible. 

The light curve was investigated using a sliding Fourier transform (cf. Fig.~\ref{f:slide}). 
The choice of the duration of each element of the sliding transform is a compromise  between temporal and frequency resolution; samples of duration 10\,d giving a frequency resolution of $\approx 0.1\perday$ were found to give the most coherent picture. 
As for the FT of the complete dataset, no single picture emerges. 
A strong signal at $f\approx0.4\perday$ is seen between {\it Kepler} days 2480 and 2500 and also around day 2540.  
Another strong signal at $f\approx0.1-0.2\perday$ is seen from days 2520 to 2545.  
Higher frequency signals are sporadic and not coherent. 

 {\it K2} observed up to 15 cool RCB stars during Campaigns 7, 9 and 11 \citep{clayton17}. Inspection of the {\it Kepler} database shows clear evidence of cyclical variation on a time scale of $\approx 40\,{\rm d}$ for several stars, whilst others were observed during their slow rise to maximum light. Analyses are continuing.

\section{Pulsation and Dust }

\subsection{Pulsation}

Deducing a physical mechanism for incoherent light variations is difficult. 
Light variations of similar amplitude and timescale have been observed in other low-mass hydrogen-deficient supergiants of similar effective temperature, namely the EHe stars V2076\,Oph \citep{lynasgray87} and V2209\,Oph \citep{jeffery85b}. 
Whilst light curves of both stars have been interpreted in terms of a multi-periodic variation, their duration and coverage make such a conclusion risky \citep{wright06}. 
Nevertheless, the envelopes of hydrogen-deficient supergiants are unstable to pulsation through opacity-driven strange-mode instability on approximately the timescales observed \citep{saio88b,jeffery16a}. 
Consequently, a legitimate but not exclusive interpretation of the {\it K2} lightcurve is a quasiperiodic oscillation of the stellar photosphere with periods of $\approx 2.5$ and 5 -- 10\,d. 
These are comparable with those seen in V2209\,Oph \citep{jeffery85b}. 
The incoherent flickering seen at 0.5\,d timescales does not fit the pulsation picture so well. 

\subsection{Dust triggers}

A corollary to the variability of V348\,Sgr at maximum light is provided by cooler RCB variables on longer timescales. 
Early observations of RY\,Sgr attempted to identify a correlation between pulsation phase and decline onset \citep{alexander72}. 
Further efforts to establish pulsation characteristics have recognised that the light (and velocity) variations in cool RCB stars are not strictly periodic \citep{crause07}.
Moreover, the locale of dust formation is restricted to a relatively small fraction of the surface \citep{jeffers12,bright11}.
Synoptic spectroscopy of R\,CrB itself has provided evidence for increased turbulent activity in the stellar atmosphere prior to onset of the great decline of 2007 \citep{feast19}.  

A physical model has been proposed in which carbon nucleation could be triggered by accelerated cooling following the periodic passage of an outward-running shock wave \citep{woitke96}. 
This model provides a crucial link between the phenomena of dust formation and pulsation in cool RCB stars on the one hand, and evidence that the dust must form at small radial distance from the star, on the other \citep{clayton96}.
It is conditional on the nucleation growth time not exceeding the pulsation period. 
Whilst there is good evidence for a pulsation-decline relation in RCB stars \citep{crause07}, it still remains to be demonstrated  that a dust episode is triggered by a pulsational excursion to maximum radius and not, for example, by the eruption of a large convection cell as suggested by \citet{wdowiak75} (see also \citet{feast86,feast96}). 
That the question is still live is attested by \citet{feast19} who remark that {\it Coravel} data ``show that a major ‘disturbance’ was present in the atmosphere of R\,CrB for at least 130 days prior to the 2007 RCB-type decline''.

Whilst irregular or quasi-periodic pulsations may offer a plausible explanation for the light variations observed in V348\,Sgr, 
the extension of the cool-RCB dust-formation paradigm to V348\,Sgr causes difficulties. 
The photosphere of V348\,Sgr is some three times hotter than that of the cool RCB stars, and some 19\,000\,K hotter than the gas temperature $T_{\rm g}\leq1500$\,K required for carbon nucleation. 
From a simple energy density argument and assuming the luminosities of RCB stars and V348\,Sgr are similar ({\it i.e.} $T_{\rm g}\propto r^{-0.5}$, where $r$ is radial distance), the equivalent nucleation radius for V348\,Sgr would be $\approx18\Rstar$ (108\Rsolar) provided that superadiabatic cooling occurs.
Without extra cooling, $T_{\rm g}\leq1\,500$\,K would occur at $r>200\Rstar$ (1200\Rsolar). 
Moreover, this must occur over a much larger solid angle around the star in order to provide the same obscuration for a larger fraction of time. 
A second problem is that the {\it K2} ``pulsation'' period of $\approx 2.5$\,d and the 0.5\,d flickering are short compared with published observations of decline to minimum \citep[4 -- 6\,d;][]{heck81,heck82}. 
The longer {\it K2} period of 5 -- 10 d might be more accommodating, and continuing ROAD observations  may show a decline time shorter than this.
However, it is difficult to see how either pulsation or convection induced shocks (in a radiative photosphere) could trigger dust formation
at  $r>18\Rstar$. 
A possible solution lies in the presence of a stellar wind sufficiently dense to produce the coolest [WC]-type spectrum yet known. 
Could it be possible that instabilities in the wind trigger turbulence accompanied by shocks which allow multiple regions to condense superadiabatically after they pass? 
Could the {\it K2} lightcurve be associated with instabilities in the wind, rather than in the stellar photosphere?  
Modelling such phenomena has not yet been attempted. 

\begin{figure*}
\begin{center}
    \epsfig{file=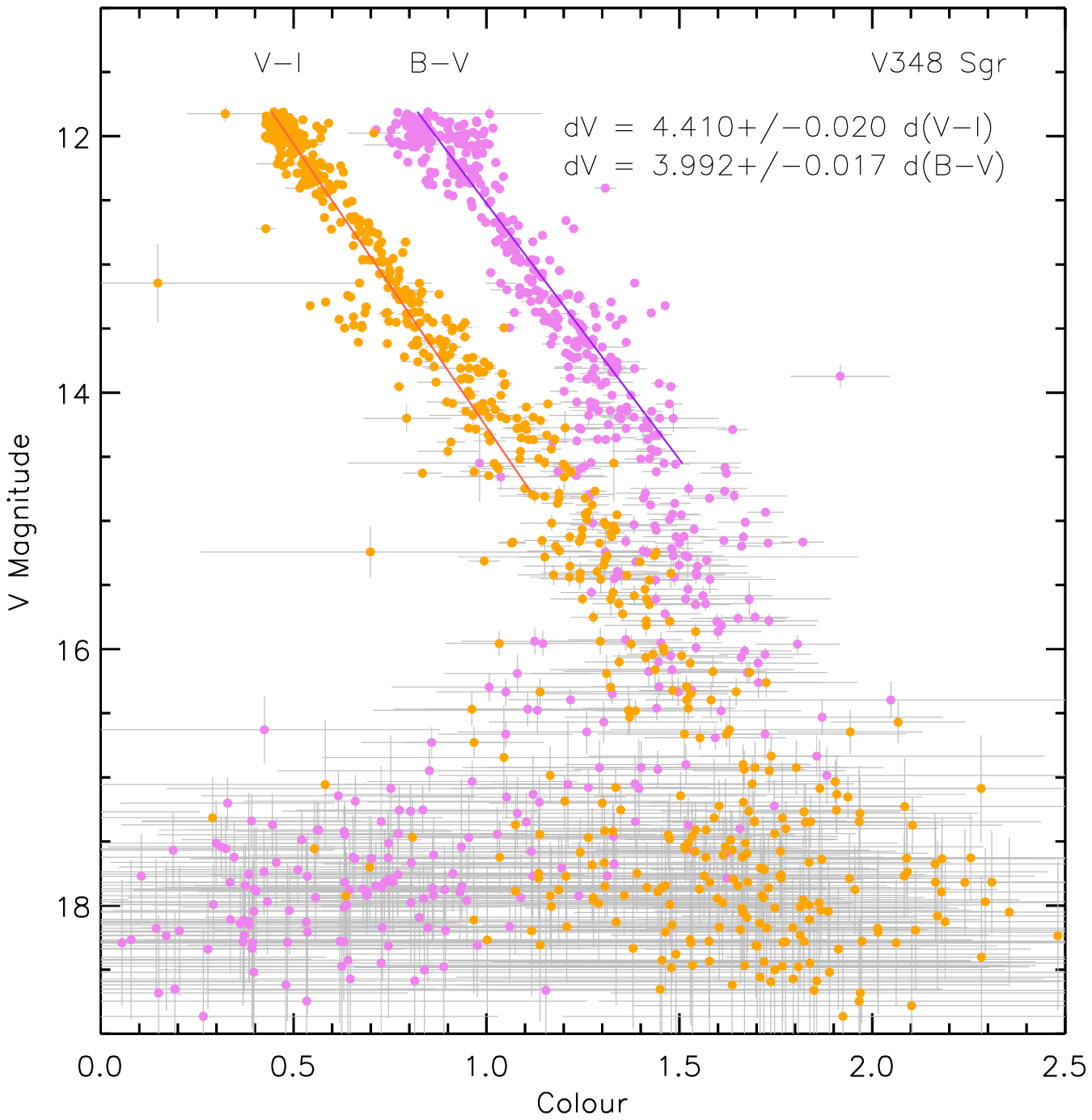,width=80mm,clip=}
    \epsfig{file=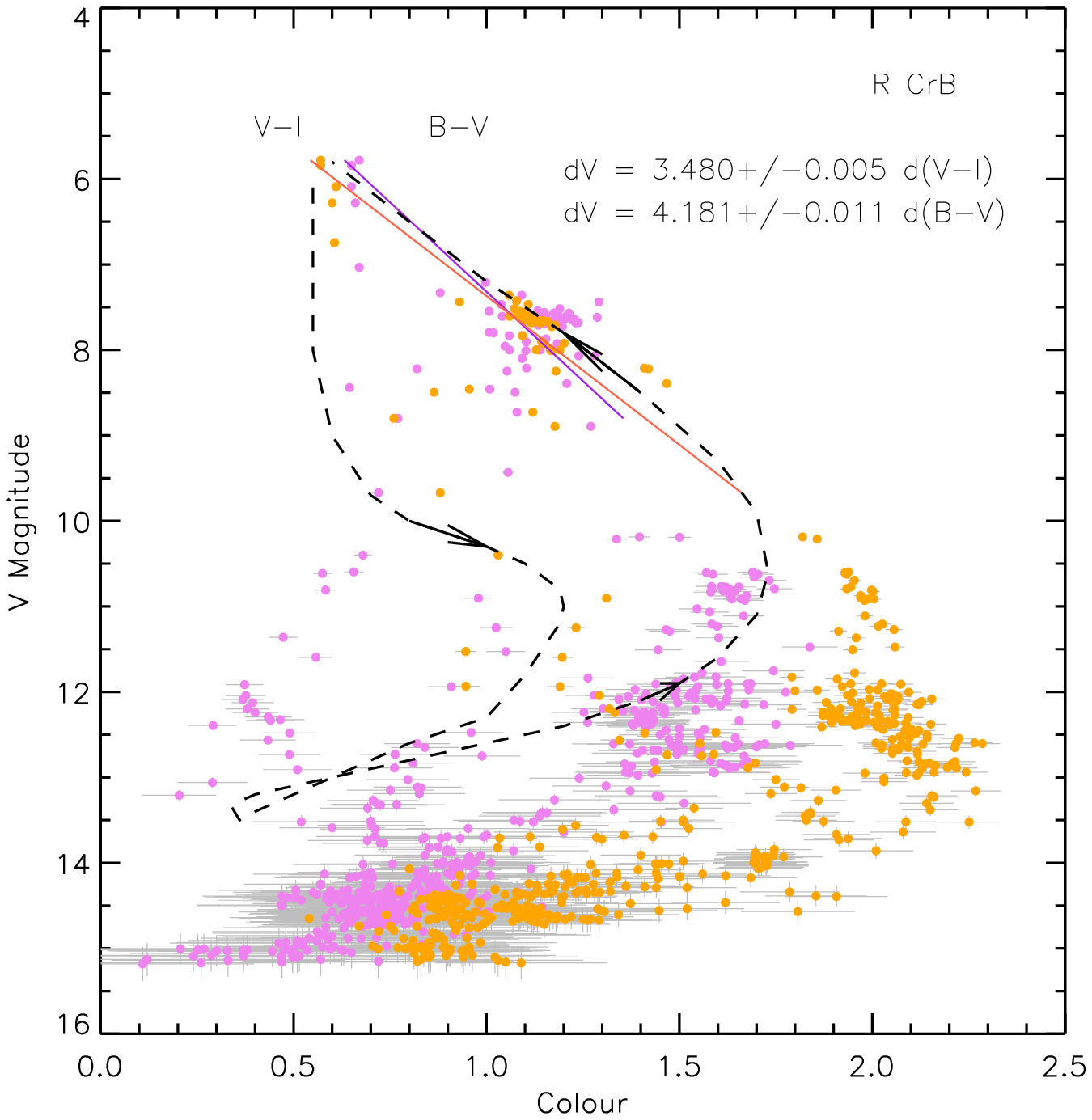,width=80mm,clip=}
\end{center}
\caption{Colour-magnitude diagrams for V348\,Sgr (left) and R\,CrB (right) based on photometry presented in this paper (V348\,Sgr) and obtained from the AAVSO (R\,CrB), using all epochs shown in Fig.\,\ref{f:lc}. Straight lines represent the fits described in \S\,4. 
The broken line in the R\,CrB panel represents the extinction track from Fig.\,5 of \citet{pugach91}.} 
\label{f:colmag}
\end{figure*}

\subsection{Dust properties}
\label{s:dust}

It is commonly understood that RCB extinction events lead to the reddening of starlight \citep{alexander72}, and hence to the inference that the extinction is caused by dust \citep{loreta35,okeefe39}. 
There is also evidence that during early stages of some declines, starlight may become bluer while the photosphere is eclipsed and the hotter corona remains visible \citep{cottrell90}. At some point during recovery, the starlight does become redder, and then recovers its original colour as the star brightens. 
Moreover, the degree of reddening during recovery carries information about the specific properties of the dust. 
The $BVI$ observations described above therefore provide an opportunity to compare the properties of dust ejected by a {\it hot} RCB star with that ejected by cool RCB stars such as R\,CrB itself. 

Figure\,\ref{f:colmag} shows the colour-magnitude ($V$ vs $B-V$ and $V-I$)  diagrams for V348\,Sgr and R\,CrB\footnote{Observations of R\,CrB obtained from the AAVSO database are described in Appendix A.}. 
Note that the zero points reflect the different intrinsic colours and non-variable components of the reddening of the two stars.
A feature hinted at by the noisy data at light minimum is observed in the ultraviolet below 3200\AA, where all of the light during a deep minimum is due to scattering  \citep{hecht98}.
At optical wavelengths, this phenomenon is more apparent in the comparatively cleaner data for R\,CrB, namely that the colours  at light minimum are similar to those at light maximum. 
In the case of V348\,Sgr, instrumental sensitivity is a limitation. 

Figure\,\ref{f:colmag} must be interpreted with caution. 
Nearly all of the observations were obtained during deep minimum, during return to maximum light or at maximum light. 
They do not indicate what happens during rapid decline. 
As stated, there is evidence that starlight becomes bluer or remains blue during this initial phase \citep{cottrell90}. 
\citet[][Fig.\,5]{pugach91}  argued that for R\,CrB there is a hysteresis in the colour-magnitude plane, with decline following a steeper and bluer trajectory than recovery. 
His numerical description for the photometric behaviour of R\,CrB is shown in Fig.\,\ref{f:colmag} in which reddening during rapid decline appears to restrict $\delta(B-V)<0.7$ mag. 
This low reddening phase may be due to  the only surviving light being scattered from small grains, rather than being absorbed and re-emitted.
Combining redder light from scattering close to the line of sight (cf. sunset) and bluer light from higher scattering angles (blue sky),
the nett effect is the reverse of the reddening seen later at smaller optical depths.
There is limited supporting evidence in the case of R\,CrB where AAVSO data points at intermediate extinctions ($12>V>9$) are seen in the blue   (Fig.\,\ref{f:colmag}), 
However, V348\,Sgr was not observed during a deep decline, and the 2007 -- 2018 sample for R\,CrB covers only 2 declines.

The evidence that V348\,Sgr commences deep minima as a blue star is most clearly seen in $B-V$ (Fig.\,\ref{f:colmag}). 
It is less obvious in $V-I$ and  possibly occurs only at higher extinctions than in $B-V$. 
R\,CrB at maximum extinction ($\delta V > 8$\,mag) has similar colours to those at light maximum. 
For both stars, as deep minimum continues, dust grains grow, the dust cloud starts to disperse and starlight becomes redder by up to 1.5 magnitudes when $\delta V \approx 6$\,mag below maximum light. 
Thereafter both stars become bluer as they recover to full brightness. 
 It is during these final stages of recovery that the dust properties of R\,CrB and V348\,Sgr can most easily be compared. 

To compare dust properties, linear fits of the form
\[
\delta V =  m_1 \delta (B-V),\,\,\,
\delta V =  m_2 \delta (V-I)
\]
were obtained by weighted regression on points with $V<15$ (V348\,Sgr) and $V<12$ (R\,CrB) and $\sigma_{B-V}<0.01$ (or $\sigma_{V-I}<0.01$).
The resulting fits are shown in Fig.\,\ref{f:colmag} and the coefficients are as follows:
\begin{tabbing}
$m_X$ \= $=$    \= $3.992\pm0.017$ ~~ \= $4.181\pm0.011$  \kill
     \>  \> V348\,Sgr ~~ \> R\,CrB  \\    
$m_1$ \> $=$    \> $3.992\pm0.017$  ~~\> $4.181\pm0.011$  \\
$m_2$ \> $=$ \> $4.410\pm0.020$  ~~\> $3.480\pm0.005$  \\
\end{tabbing}
The coefficients represent the slopes of the reddening lines obtained from the highest quality data $<3$mag below light maximum.
For R\,CrB, they can be compared with the value $m_1=5.32\pm0.31$ given as the inverse slope $1/m_1$ by \citet{pugach04}. 
The latter argued that the  extinction properties of RCB ejecta are distinct from those of dust around other stellar types ($\langle m_1 \rangle=3.98$) and in the interstellar medium, presumably because the former is  carbonaceous and the latter  silicate. However, the numbers obtained here for R\,CrB do not support the value obtained by \citet{pugach04}.
Despite the small formal errors relative to the gradient differences, sparse data coverage for R\,CrB makes any conclusion on differences between V348\,Sgr and R\,CrB premature.  

\section{Conclusion}
\label{s:conc}

V348\,Sgr continues to present observational challenges. 
New observations with ROAD and {\it K2} partially address the high fraction (45\%) of time spent in RCB-type declines.

We have discovered significant short-period ($\sim 2.5 - 10$\,d) small-amplitude (0.01\%) light variations. 
The timescale is correct for pulsations, but the origin of the activity needs to be better localized to either the photosphere or the wind. 
The timescale is short compared with historical observations of the decline phase. 

We have measured the colour-magnitude relation during deep minima and during recovery, but observations of ingress are elusive, even after five years.  
The colour-magnitude relations measured during return from light minimum strongly resemble those measured in R\,CrB. 
The latter demonstrate a gradient reversal at around 6 magnitudes of extinction in $V$. 
Taking R\,CrB and V348\,Sgr together, the {\bf obscured} spectrum appears to reach a scattering limit as the optical depth reaches unity
first in the ultraviolet \citep{hecht98}, followed by $B$, $V$, and finally $I$. 
These similarities between V348\,Sgr and R\,CrB support \citet{clayton11b} who assert that the dust properties of V348\,Sgr are more similar to cool RCB stars than to another hot RCB star (HV\,2671) and a cool Wolf-Rayet central star (CPD$-56^{\circ}8032$), and \citet{drilling97} who find the dust likely to be composed of amorphous carbon, rather than graphite.  

The fraction of time spent at minimum light also supports the idea argued by \citet{drilling97} of a cloud covering the entire star and not just lying in the line of sight. 
In light of these observations, we suggest that dust production in V348\,Sgr could occur as a result of small-scale supersonic turbulence producing localised shock fronts ubiquitously in the stellar wind at a radius of some 20\Rstar. 
Such shock fronts would lead to superadiabatic cooling and carbon nucleation on a range of length scales.  
They could also manifest as the short-characteristic (0.5\,d) incoherent flickering seen in the K2 light curve. 
Models are needed to test such ideas.

Continuing observations of V348\,Sgr are essential to probe the early phases of dust production during a deep decline. 
In view of the current impossibility  of {\it predicting} times of decline, these are difficult.  
Photometry sensitive to millimagnitude variability at maximum light and through early decline is desirable. 
Spectroscopy at similar phases would determine any lag between photospheric lines and wind lines becoming obscured, 
and hence help locate the carbon nucleation site.   
Meanwhile, broadband photometric observations are continuing at ROAD in an effort to  characterise the decline timescales.

\section*{Acknowledgments}
This paper includes data collected by the {\it Kepler} mission.  Funding for the {\it Kepler} mission is 
provided by the NASA Science Mission directorate.

Some of the data presented in this paper were obtained from the Mikulski Archive for Space Telescopes
(MAST). STScI is operated by the Association of Universities for Research in Astronomy, Inc., under NASA 
contract NAS5-26555. Support for MAST for non-HST data is provided by the NASA Office of Space
Science via grant NNX09AF08G and by other grants and contracts.

This research has made use of the SIMBAD database, operated at CDS, Strasbourg, France.

We acknowledge with thanks the variable star observations from the AAVSO International Database contributed by observers worldwide and used in this research.

CSJ thanks Churchill College Cambridge for a visiting by-fellowship and the Institute of Astronomy Cambridge for a visitor grant. 

Research at the Armagh Observatory and Planetarium is supported by a grant-in-aid 
from the Northern Ireland Department for Communities.

\bibliographystyle{mn2e}
\bibliography{ehe}

\appendix
\renewcommand\thefigure{A.\arabic{figure}} 
\renewcommand\thetable{A.\arabic{table}} 

\begin{figure*}
\begin{center}
    \epsfig{file=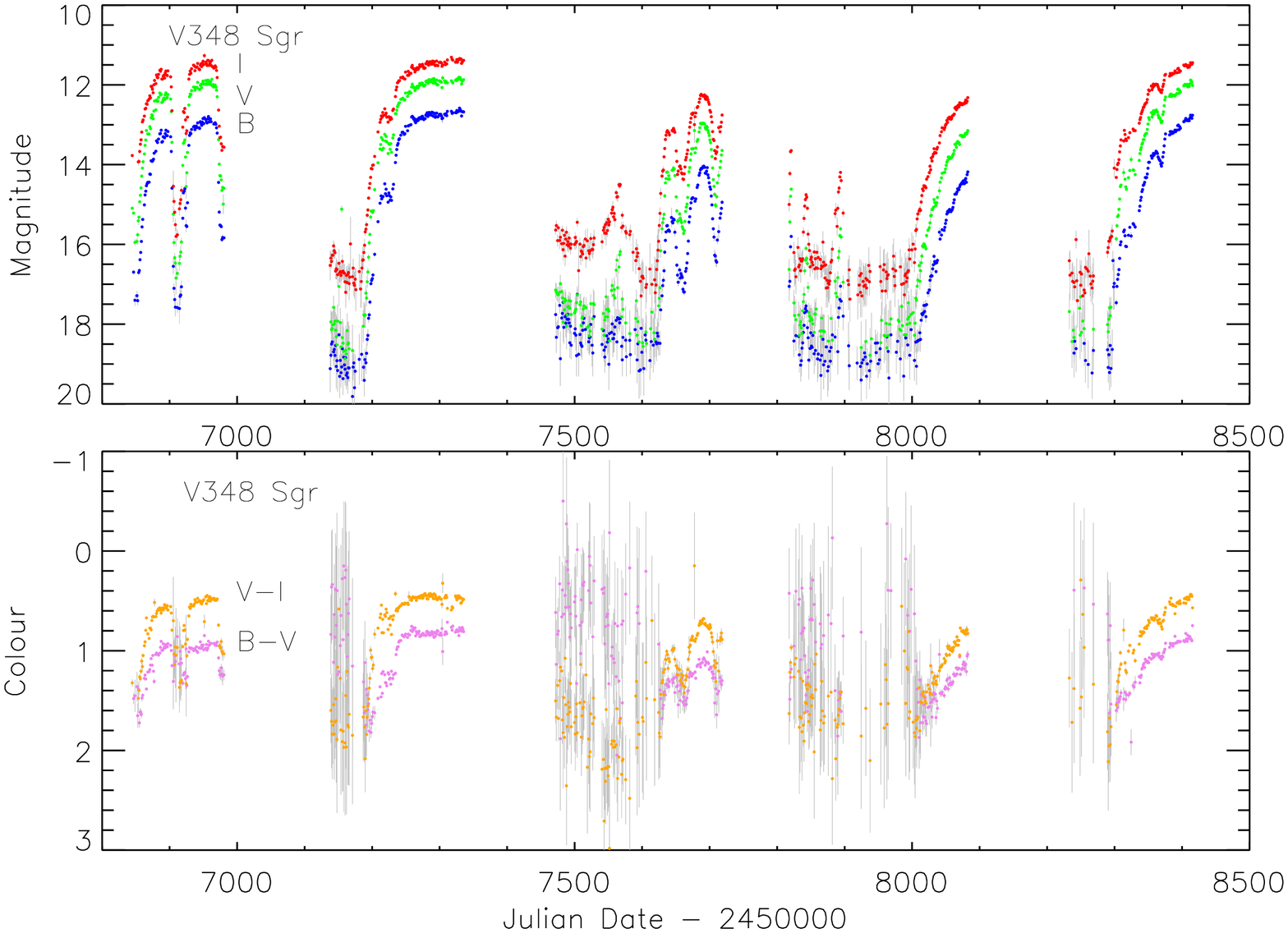,width=150mm,clip=,angle=0}
    \epsfig{file=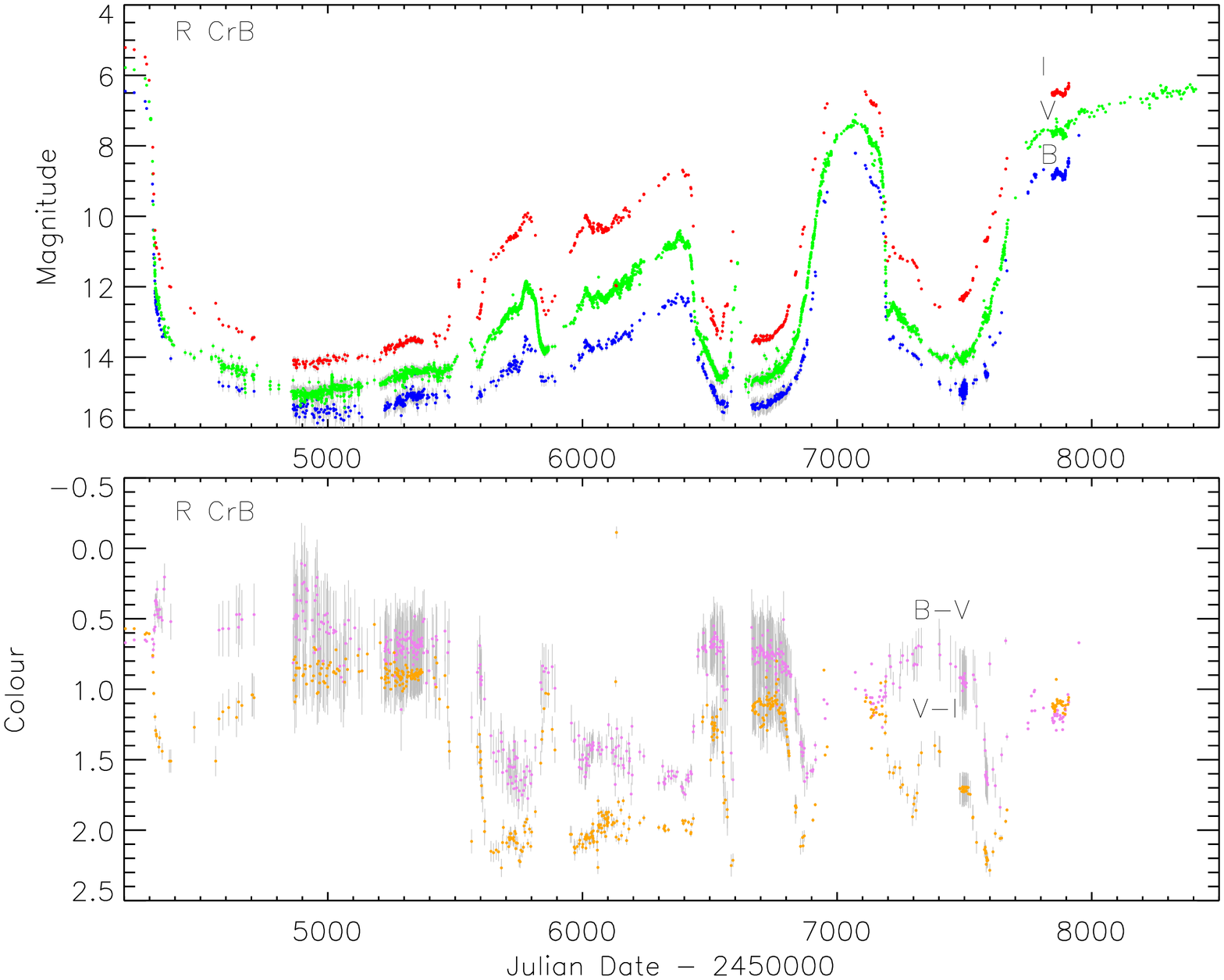,width=150mm,clip=,angle=0}
\end{center}
\caption{Three-colour ($BVI$) photometry and colours  ($B-V$ and $V-I$) of V348\,Sgr (Top pair) obtained from ROAD between 2014 and 2018 and of R\,CrB (Bottom pair) obtained from the AAVSO database and observed between 2007 and 2018. } 
\label{f:lc}
\end{figure*}

\section{AAVSO Light curves for V348~Sgr and R~CrB}
The complete V348\,Sgr light and colour curves reported in \S\,2 are shown in Fig.~\ref{f:lc}.

Photometry of R\,CrB used in this paper was obtained from the AAVSO online database and is shown in Fig.~\ref{f:lc}. The data were sifted to include only secure observations made in $B$, $V$ and $I$ filters between the dates 2007 Jan 1 and 2018 Dec 31. Upper limits were excluded. As widely reported, R\,CrB was in a low state for a substantial fraction of this interval \citep[e.g.][]{howell13}. The AAVSO observations were binned to form daily means. Colours were formed from observations obtained on the same day. Photometric errors are not reported by AAVSO for any of the data extracted here. We therefore estimated an error for every measurement $m$ (mag.) from
\[
e = \pm 10^{(m-18)/4} \,{\rm mag},
\]
corresponding to $\pm0.001$ mag at $m=6$ and $\pm0.5$ mag at $m=18$. 

The AAVSO observer codes for the data shown in Fig. \ref{f:lc} are: 
 AAM, AAUA, ACAB, ADI, ATE, BCP, BDLA, BFX, BIY, BJAA, BMAH, BOA, BPAD, BPO, BVE, CAMA, CEM, CMF, CMP, CNY, CPE, CTX, DDJ, DIL, DJED, DKS, DUBF, EJC, FJQ, FRL, GCJ, GCO, GHN,
GOT, GTZ, GXR, HBB, HGUA, HPIA, HSR, HTY, HUR, HUZ, JAZ, JSJA, KIR, KJGB, KMM, KSQ, KTHA, KVI, LAL, LCLA, LMJ, MAV, MDJ, MDW, MEV, MFB, MQE, MRV, MUY, MXL, MZK, NKL, NLZ, NMR,
NOT, OALA, OCN, ONJ, PAGA, PCG, PDAE, PGU, PKV, PLN, PNIB, PNQ, PWD, PXR, PYG, RGN, RJV, RMN, ROE, RZD, SAND, SBIA, SC, SDI, SDN, SFRA, SGOR, SHA, SHS, SJAR, SRIC, SSTB, STAC,
SXN, TIA, TSZ, UIS01, UMAA, VOL, VST, VWA, WDO, WDZ, WFOA, WKL, WTHB and ZIN. 
Individual observers may be identified at the AAVSO website: {\tt  https://www.aavso.org}.

\label{lastpage}
\end{document}